\documentclass[jap,twocolumn,a4paper,showpacs]{revtex4-1}

\usepackage[utf8]{inputenc}
\usepackage{amsmath,amsfonts,amssymb}
\usepackage{graphicx,graphics}
\usepackage{color,textcomp,bm}
\usepackage{subfigure,array,xspace} 
\usepackage[squaren,Gray,cdot]{SIunits}

\usepackage[pdfstartview=FitV,bookmarks=true,
            linktocpage=true,colorlinks=true,
            urlcolor=blue,linkcolor=red, 
            citecolor=green]{hyperref}
            
\newcommand{\ie}{\textit{i.e.}\@\xspace}
\newcommand{\eg}{\textit{e.g.}\@\xspace}

\newcommand{\etal}{\emph{et al.~}}

\begin{document}

\title{Investigation of interfacial stiffnesses of a tri-layer\\ using Zero-Group Velocity Lamb modes}

\author{Sylvain Mezil}      
\author{Francois Bruno}     
\author{Samuel Raetz}       
\author{J\'er\^ome Laurent} 
\author{Daniel Royer}       
\author{Claire Prada}       
\affiliation{ESPCI ParisTech, PSL Research University, CNRS, Institut Langevin, 1 rue Jussieu, F-75005, Paris, France}

\begin{abstract}
Zero-Group Velocity (ZGV) Lamb waves are studied in a structure composed of two plates bonded by an adhesive layer. The dispersion curves are calculated for a Duralumin/epoxy/Duralumin sample, where the adhesion is modeled by a normal and a tangential spring at both interfaces. Several ZGV modes are identified and their frequency dependence on interfacial stiffnesses and on the bonding layer thickness are numerically studied. Then, experiments achieved with laser ultrasonic techniques are presented. Local resonances are measured using superimposed source and probe. Knowing the thicknesses and elastic constants of the Duralumin and epoxy layers, the comparison between theoretical and experimental ZGV resonances leads to an evaluation of the interfacial stiffnesses. A good agreement with theoretical dispersion curves confirms the identification of the resonances and the parameter estimations. This non-contact technique is promising for the local evaluation of bonded structures. 
\end{abstract}


\maketitle

\section{Introduction}\label{sec:intro}

The characterization of the elastic coupling of two plates by an adhesive layer is of primary importance in the evaluation of bonded structures. Various acoustic methods have been proposed to measure adhesion using reflection~\cite{article-rokhlin-ultrasonic} or transmission~\cite{article-budzik} of bulk acoustic waves. These methods are more suitable for disbond detection than for adhesion characterization. Other techniques exploit guided waves, such as Lamb modes or shear horizontal modes, propagating along the structure.~\cite{article-crom, article-vlasie-acoustical,article-vlasie-ultrasonic, article-alaoui,article-cerniglia} However, these non local methods are not very sensitive and provide an average estimation of the bonding quality. A thorough review on ultrasound non-destructive testing of adhesive bond was conducted by Castaings.\cite{article-castaings}

An alternative approach using Zero-Group Velocity (ZGV) Lamb modes has been the object of recent studies. The existence of backward waves, and consequently of ZGV modes, was studied theoretically for solid/liquid/solid structures.~\cite{article-nishimiya, article-nishimiya-relationships, article-mezil-bci} Laser ultrasonic experiments confirmed the narrow resonances observed at ZGV frequencies in the case of a thin fluid layer between two solid plates.~\cite{article-mezil-bci} It was also shown experimentally that ZGV modes exist in plates bonded by a solid layer and are promising to evaluate the bonding quality.~\cite{article-prada-laser-ultrasonic, article-matsuo-damage, article-matsuo-evaluation} High sensitivity and local measurements provided by ZGV modes, are attractive for investigating the bonding in a tri-layer structure without any contact. Using resonance frequencies of ZGV Lamb modes, Cho \etal reported on the evaluation of well- and weak-bonded adhesive plate samples.~\cite{article-matsuo-evaluation, article-cho-characterization} They observed a correlation between the first ZGV frequency and the bonding quality measured by mechanical tests.\\

In this article, we investigate the evolution of the main ZGV modes in a tri-layer structure as a function of the longitudinal and shear interfacial stiffnesses. In the first section, the dispersion equation of a symmetrical tri-layered structure with interfacial stiffnesses is recalled and ZGV modes are discussed. The second section propose a numerical study of the dependence of ZGV modes on the bonding layer thickness and on the interfacial stiffnesses. Then experimental results obtained with a laser ultrasonic set-up are presented. 

\section{\label{sec:disp_curves} Dispersion equations and ZGV modes of a tri-layer}

\begin{figure}[ht!]
\centering
\includegraphics[width=0.9\columnwidth]{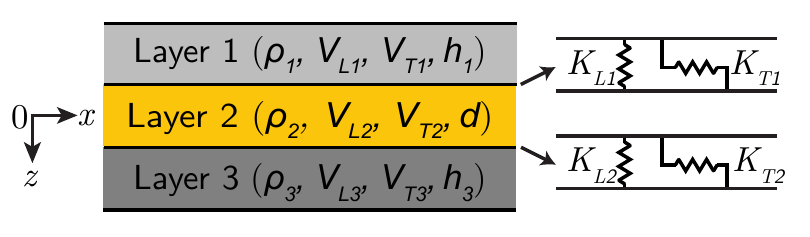}
\caption{Geometry of the tri-layer model.}
\label{fig:schema1}
\end{figure}

The objective of this section is to establish the dispersion equation of a tri-layered structure following the approach chosen by Vlasie and Rousseau.~\citep{article-vlasie-acoustical} Each layer is assumed to be isotropic, homogeneous and infinite, with mass density $\rho_i$, longitudinal and transversal velocities $V_{Li}$ and $V_{Ti}$, where $i$ indicates the layer number ($i=1,2,3$). The first and third layer thicknesses are denoted by $h_1$ and $h_3$, whereas the bonding layer thickness is denoted $d$. The coupling between two successive layers is modeled by longitudinal and transversal springs of stiffnesses $K_{Lj}$ and $K_{Tj}$, where $j=1,2$ indicates the interface [Fig.~\ref{fig:schema1}]. Spring stiffnesses can evolve from 0 (for uncoupled plates) up to $10^{17}$~N/m$^{3}$. The upper limit corresponds to an estimation, in an atomic chain model, of the epoxy layer bulk stiffnesses per unit length (N/m$^3$ or Pa/m).~\cite{livre-royer-ang}

In order to establish the dispersion law between the angular frequency $\omega$ and the wave number $k$, the scalar potential $\phi$ and the vector potential $\psi$ are used. The problem is supposed two dimensional so that the vector potential reduces to a scalar. The tangential and normal displacements derive from these potentials as follows:
\begin{equation}
u_{x}=\frac{\partial\phi}{\partial x}	-\frac{\partial\psi}{\partial x}\text{, }\quad u_{z}=\frac{\partial\phi}{\partial x}+\frac{\partial\psi}{\partial x}
\label{eq:ux_uz}
\end{equation}
and the stresses are given by:
\begin{eqnarray}
\sigma_{xz}&=&\mu\left(\frac{2\partial^2\phi}{\partial x\partial z}+\frac{\partial^2\psi}{\partial x^2}-\frac{\partial^2\psi}{\partial z^2}\right),\\
\sigma_{zz}&=&\lambda\left(\frac{\partial^2\phi}{\partial x^2}+\frac{\partial^2\phi}{\partial z^2}\right)+2\mu\left(\frac{\partial^2\phi}{\partial z^2}+\frac{\partial^2\psi}{\partial x\partial z}\right),
 \end{eqnarray}
where $\lambda$, $\mu$ are the Lam\'e coefficients.~\cite{Achenbach73} 
The potentials in the three layers are written as:
\begin{widetext}
\begin{equation}
\begin{cases}
\phi_{1}\!=\!\left(A_{1L}\cos\!\left[p_{1} \left(z+\frac{d}{2}\right)\right]\!+\!B_{1L}\sin\!\left[p_{1} \left(z+\frac{d}{2}\right)\right]\right)e^{\jmath\left(k x-\omega t\right)},\\
\psi_{1}\!=\!\left(A_{1T}\cos\!\left[q_{1} \left(z+\frac{d}{2}\right)\right]\!+\!B_{1T}\sin\!\left[q_{1} \left(z+\frac{d}{2}\right)\right]\right)e^{\jmath\left(k x-\omega t\right)},\\
\phi_{2}\!=\!\left(A_{2L}\cos\!\left[p_{2} z\right]\!+\!B_{2L}\sin\!\left[p_{2} z\right]\right)e^{\jmath\left(k x-\omega t\right)},\\
\psi_{2}\!=\!\left(A_{2T}\cos\!\left[q_{2} z\right]\!+\!B_{2T}\sin\!\left[q_{2} z\right]\right)e^{\jmath\left(k x-\omega t\right)},\\
\phi_{3}\!=\!\left(A_{3L}\cos\!\left[p_{3} \left(z-\frac{d}{2}\right)\right]\!+\!B_{3L}\sin\!\left[p_{3} \left(z-\frac{d}{2}\right)\right]\right)e^{\jmath\left(k x-\omega t\right)},\\
\psi_{3}\!=\!\left(A_{3T}\cos\!\left[q_{4} \left(z-\frac{d}{2}\right)\right]\!+\!B_{3T}\sin\!\left[q_{3} \left(z-\frac{d}{2}\right)\right]\right)e^{\jmath\left(k x-\omega t\right)},
\end{cases}
\label{eq:psiphi}
\end{equation}
\end{widetext}
where $p$ and $q$ are the {\it z}-component of the longitudinal and transversal wave vectors, respectively. The wave numbers $k_{Li}=\omega/V_{Li}$ and $k_{Ti}=\omega/V_{Ti}$ satisfy dispersion equations of bulk waves ${k_{Li}}^2={k}^2+{p_i}^2$ and ${k_{Ti}}^2={k}^2+{q_i}^2$. $A_{iL}$ and $B_{iL}$ ($A_{iT}$ and $B_{iT}$) are amplitudes of longitudinal (shear) components. At the free boundary, tangential ($\sigma_{xz}$) and normal ($\sigma_{zz}$) stresses vanish:
\begin{equation}
\begin{cases}
\sigma_{zz1}=\sigma_{xz1}=0 & \text{for $z=-h_1-d/2$},\\
\sigma_{zz3}=\sigma_{xz3}=0 & \text{for $z=h_3+d/2$}.
\label{eq:condlim}
\end{cases}
\end{equation}
The spring boundary conditions at both epoxy interfaces~\cite{article-jones-whittier} are written as follows:
\begin{equation}
\begin{cases}
\sigma_{zz1}=\sigma_{zz2}=K_{L1}\left(u_{z2}-u_{z1}\right) &  \text{for $z=-d/2$},\\
\sigma_{xz1}=\sigma_{xz2}=K_{T1}\left(u_{x2}-u_{x1}\right) & \text{for $z=-d/2$},\\
\sigma_{zz2}=\sigma_{zz3}=K_{L2}\left(u_{z3}-u_{z2}\right) &  \text{for $z=d/2$},\\
\sigma_{xz2}=\sigma_{xz3}=K_{T2}\left(u_{x3}-u_{x2}\right) & \text{for $z=d/2$}.
\label{eq:condlim}
\end{cases}
\end{equation}
Assuming a symmetric structure, the first and third layers properties, as well as the two longitudinal and normal springs are identical and denoted $h$, $\rho_1$, $V_{L1}$, $V_{T1}$, $K_{L}$ and $K_T$, respectively. Then, it can be demonstrated that the system can be split into two sub-matrices, corresponding to symmetrical and anti-symmetrical solutions. Symmetrical modes fulfill the conditions:
\begin{equation}
\begin{cases}
u_x(z)=u_x(-z),\\
u_z(z)=-u_z(-z),
\label{eq:cas_symetrique}
\end{cases}
\end{equation} 
whereas anti-symmetrical modes satisfy:
\begin{equation}
\begin{cases}
u_x(z)=-u_x(z),\\
u_z(z)=u_z(-z).
\label{eq:cas_antisymetrique}
\end{cases}
\end{equation}
Combining Eqs.~(\ref{eq:ux_uz}, \ref{eq:psiphi}, \ref{eq:cas_symetrique}, \ref{eq:cas_antisymetrique}) leads to:
$A_{1L}=A_{3L}$, $B_{1T} =B_{3 T}$, $A_{1T}=-A_{3T}$, $B_{1L} =-B_{3L}$ and $A_{2T}=B_{2L}=0$ for symmetrical modes. And to: $A_{1L}=-A_{3L}$, $B_{1T} =-B_{3 T}$, $A_{1T}=A_{3T}$, $B_{1L} =B_{3L}$ and $A_{2L}=B_{2T}=0$ for anti-symmetrical modes. Introducing the matrix:\\ 
\begin{widetext}
\centering
\begin{displaymath}
M(\alpha)=\left[ \begin{array}{cccc}
2 \jmath k p_1 \sin[p_1 h] & 2 \jmath k p_1 \cos[p_1 h] & ({k_{t1}}^2-2 k^2) \cos[q_1 h] & -({k_{t1}}^2-2 k^2) \sin[q_1 h]  \\
-({k_{t1}}^2-2 k^2) \cos[p_1 h] & ({k_{t1}}^2-2 k^2) \sin[p_1 h] & 2 \jmath k q_1 \sin[q_1 h] & 2 \jmath k q_1 \cos[q_1 h] \\
\jmath k K_T & 2 \jmath k p_1 \mu_1 & ({k_{t1}}^2-2 k^2) \mu_1 & -q_1 K_T\\
-({k_{t1}}^2-2 k^2) \mu_1 & p_1 K_L & \jmath k K_L & 2 \jmath \mu_1 k q_1\\
\jmath k K_T & 0 & 0 & -q_1 K_T \\
 0 & p_1 K_L & \jmath k K_L & 0
\end{array} \right.
\end{displaymath}
\centering
\begin{multline}
\left. \begin{array}{cc}
 0                                         & 0\\
 0                                         & 0\\
-\jmath k K_T \cos[p_2 \frac{d}{2}+\alpha] & q_2 K_T \cos[q_2 \frac{d}{2}-\alpha]\\
-p_2 K_L \sin[p_2 \frac{d}{2}+\alpha]      & \jmath k K_L \sin[q_2 \frac{d}{2}-\alpha]\\
2 \jmath k p_2 \mu_2 \sin[p_2 \frac{d}{2}+\alpha]-\jmath k K_T \cos[p_2 \frac{d}{2}+\alpha] & -({k_{t2}}^2-2 k^2) \mu_2 \sin[q_2 \frac{d}{2}-\alpha]+q_2 K_T \cos[q_2 \frac{d}{2}-\alpha] \\
-({k_{t2}}^2-2 k^2) \mu_2 \cos[p_2 \frac{d}{2}+\alpha]-p_2 K_L \sin[p_2 \frac{d}{2}+\alpha] & 2 \jmath k q_2 \mu_2 \cos[q_2 \frac{d}{2}-\alpha]+\jmath k K_L \sin[q_2 \frac{d}{2}-\alpha]\end{array} \right],
\label{eq:matriceS}
\end{multline}
\end{widetext}
the boundary conditions can be written in matrix form by setting $\alpha=0$ for symmetrical modes and $\alpha=\pi/2$ for anti-symmetrical modes. It comes:
\begin{eqnarray*}
M_S U_S=0 \text{ where } 
U_S &=& \begin{bmatrix} A_{1L} \\ B_{1L} \\ A_{1T} \\ B_{1T} \\  A_{2L} \\ B_{2T} \end{bmatrix}  
\text{ and } M_S=M(\alpha=0),
\end{eqnarray*}
\begin{eqnarray*}
M_A U_A=0 \text{ where }
U_A &=& \begin{bmatrix} A_{1L} \\ B_{1L} \\ A_{1T} \\ B_{1T} \\  B_{2L} \\ A_{2T} \end{bmatrix}
\text{ and } M_A=M(\alpha=\pi/2). 
\label{eq:potential}
\end{eqnarray*}
Non trivial solutions are found when the determinant of matrices $M_S$ or $M_A$ vanishes. For symmetrical modes, $p_1$, $q_1$ and $q_2$ can be factorized in $2^\text{nd}$, $4^\text{th}$ and $6^\text{th}$ rows, respectively, in matrix $M_S$. 
Similarly, $p_1$, $q_1$ and $p_2$ can be factorized for $2^\text{nd}$, $4^\text{th}$ and $5^\text{th}$ row in matrix $M_A$. These terms correspond to bulk waves propagating at velocities $V_{Li}$ ($p_i=0$) and $V_{Ti}$ ($q_i=0$). Such factorization avoids the unwanted bulk modes in the dispersion curves.\\

\begin{figure}[b!]
\centering
\subfigure{\includegraphics[width=\columnwidth]{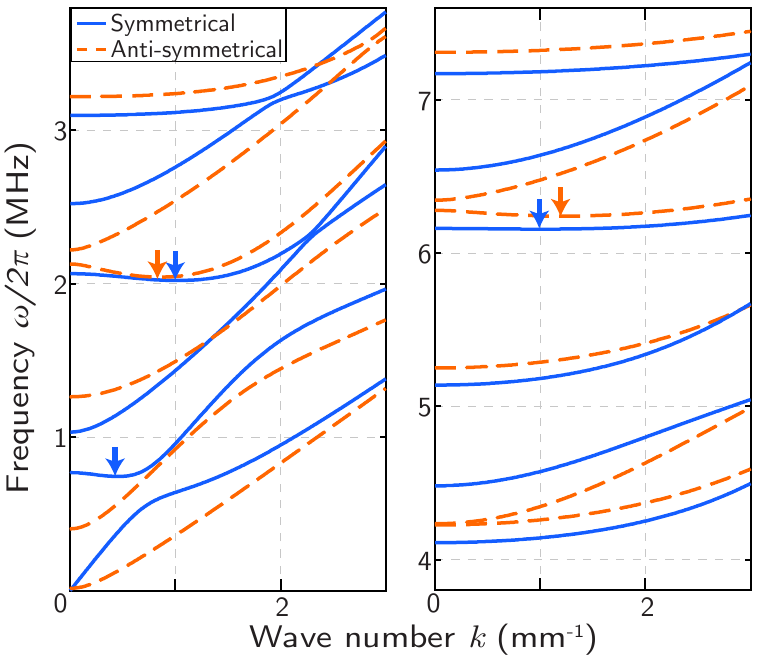}\label{figa:dispcurve_KLKT_1e5}}
\subfigure{\label{figb:dispcurve_KLKT_1e5}}
\caption{Theoretical dispersion curves for a Duralumin/epoxy/Duralumin sample, with $K_L=K_T=5\times10^{14}$~N/m$^3$ and $h=50$~\textmu m. Arrows indicate ZGV Lamb mode frequencies.}
\label{fig:dispcurve_KLKT_1e5}
\end{figure}

A numerical program was developed to calculate the dispersion curves of any isotropic tri-layer structures. Zeros of the secular equation are determined and the different modes are then distinguished. Once a mode is obtained, its phase velocity ($V_\phi=\omega/k$) and its group velocity ($V_g=\partial\omega/\partial k$) are calculated. A solution $(\omega, k)$ is a ZGV mode if $V_g=0$ with $k\neq0$. Dispersion curves and zero-group velocity Lamb waves are then obtained for the tri-layer model using interfacial stiffnesses. Furthermore, once the dispersion curves are calculated, normal and tangential displacements ($u_z$ and $u_x$, respectively) can also be obtained. For a given $(\omega, k)$ belonging to one branch of the dispersion curves, by setting one component common to $U_{S}$ and $U_{A}$ (\eg, $A_{1L}=1$), the other components are determined from the boundary equation $[M][U]=[0]$. In the general case, i.e., if layers 1 and 3 are not identical, resolution of Eqs.~\ref{eq:psiphi} and \ref{eq:condlim} leads to a 12$\times$12 matrix which cannot be separated into two submatrices. Symmetrical and anti-symmetrical modes no longer exist. In the presented example, the symmetrical tri-layer is composed of two Duralumin plates ($\rho_1=2795$~kg/m$^3$, $V_{L1}=6398$~m/s, $V_{T1}=3122$~m/s, $h=1.5$~mm) bonded with an epoxy layer ($\rho_2=1170$~kg/m$^3$, $V_{L2}=2500$~m/s, $V_{T2}=1100$~m/s, $d=50$~\textmu m). The chosen parameters correspond to measured values.\\

Figure~\ref{fig:dispcurve_KLKT_1e5} displays theoretical dispersion curves derived from Eq.~\ref{eq:matriceS} for values $K_L=K_T=5\times10^{14}$~N/m$^3$ corresponding to a relatively weak coupling. The first three branches correspond to $A_0$, $S_0$ and $A_1$ Lamb modes and like in a single plate do not present any ZGV mode. Five ZGV modes are visible on higher order branches. They can be ascribed to the repulsion of two modes.~\cite{article-prada-local} These interactions only occurs for modes of the same symmetry having close enough cut-off frequencies.\\

\begin{figure}[b]
\centering
\subfigure{\includegraphics[width=\columnwidth]{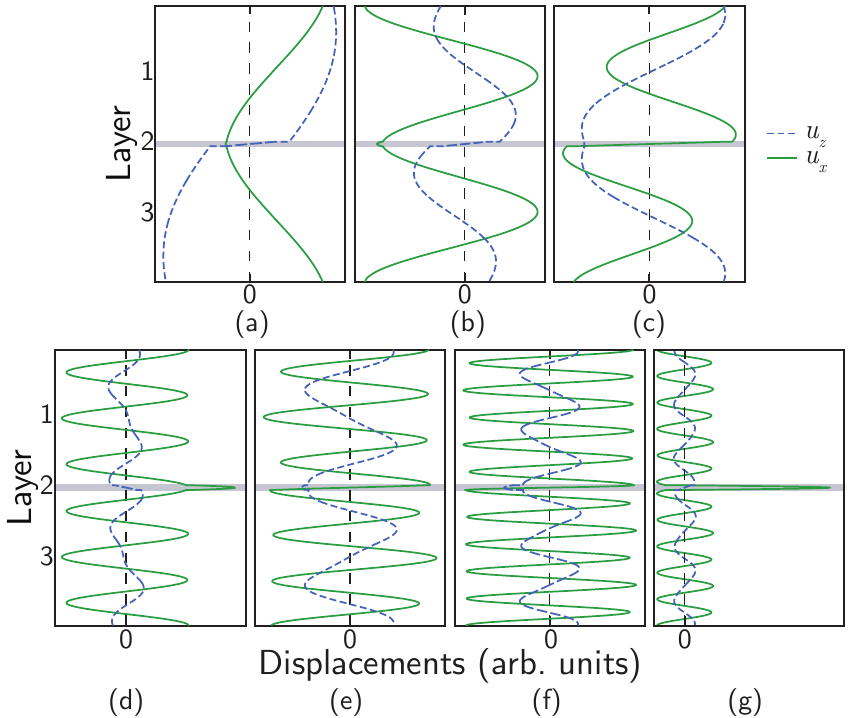}\label{figa:displacements}} 
\subfigure{\label{figb:displacements}} 
\subfigure{\label{figc:displacements}}
\subfigure{\label{figd:displacements}} 
\subfigure{\label{fige:displacements}}
\subfigure{\label{figf:displacements}} 
\subfigure{\label{figg:displacements}}
\caption{Theoretical normal (dashed line) and in-plane (solid line) displacements, in a Duralumin/epoxy/Duralumin structure with $K_L=K_T=5\times10^{14}$~N/m$^3$ and $h=50$~\textmu m, for ZGV Lamb modes (a)~${S_1S_2}$ (0.744~MHz), (b)~${S(S_1S_2)}$ (2.022~MHz), (c)~${A(S_1S_2)}$ (2.032~MHz), (d)~${S(S_3S_6)^*}$ (6.173~MHz), (e)~${A(S_3S_6)^*}$ (6.222~MHz), (f)~${A(S_5S_{10})^*}$ (10.403~MHz) and (g)~~${S(S_5S_{10})^*}$ (10.685~MHz).}
\label{fig:displacements}
\end{figure}

\begin{figure*}[ht!]
\centering
\begin{minipage}[c]{0.7\textwidth}
\subfigure{\includegraphics[width=0.9\textwidth]{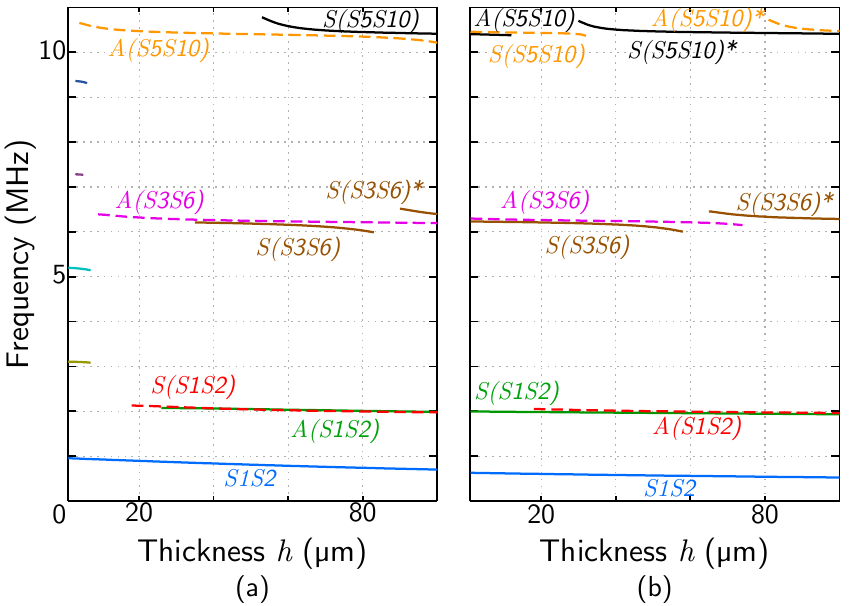}\label{figa:EvolZGV_d_KLKT_1e5}}
\subfigure{\label{figb:EvolZGV_d_KLKT_1e5}}
\end{minipage}\hfill
\begin{minipage}[c]{0.25\textwidth}
\caption{Theoretical ZGV Lamb mode frequencies as a function of the bonding layer thickness $d$, for a Duralumin/epoxy/Duralumin tri-layer, with (a)~$K_L=K_T=10^{17}$~N/m$^3$ and (b)~~$K_L=K_T=10^{14}$~N/m$^3$. Symmetrical modes are in solid lines whereas anti-symmetrical modes in dashed lines.}
\label{fig:EvolZGV_d_KLKT_1e5}
\end{minipage}
\end{figure*}

The number of ZGV Lamb modes can then vary with the different parameters. This phenomenon is discussed in Sec.~\ref{sec:params} [Figs.~\ref{fig:EvolZGV_d_KLKT_1e5}, \ref{figa:EvolZGV_KLKT_1e1_1e5}, and \ref{fig:EvolZGV_KL_fixe_KT_evol_KT_fixe_KL_evol}]. The first ZGV Lamb mode at frequency $f = 0.744$~MHz only exists in a tri-layer plate. According to the displacement distribution, this mode corresponds to the $S_1S_2$ ZGV mode of the whole structure~[Figs.~\ref{figa:displacements}], and is denoted as so hereafter. Around 2, 6 and 10~MHz, ZGV Lamb modes can be associated in pairs [see Fig.~\ref{fig:dispcurve_KLKT_1e5} for the pairs around 2 and 6~MHz]. These pairs correspond to symmetrical and anti-symmetrical coupling of same ZGV modes of the single plate. Figure~\ref{fig:displacements} displays the normal and shear displacements of the first seven ZGV Lamb modes for the same intermediate bonding ($K_L=K_T=5\times10^{14}$~N/m$^3$).\\

From Figs.~\ref{figb:displacements}-\ref{figc:displacements}, it appears that the two modes around 2~MHz correspond to the ZGV Lamb mode ${S_1S_2}$ in a single Duralumin plate where both plates are coupled by the intermediate layer symmetrically and anti-symmetrically, respectively. Their frequencies remain close: 2.022 and 2.032~MHz to be compared with 1.915~MHz for a single 1.5-mm thick Duralumin plate. The corresponding modes are denoted ${S(S_1S_2)}$ and ${A(S_1S_2)}$ hereafter. Similar observation is possible for higher frequency modes display in Fig.~\ref{fig:displacements}: ZGV Lamb modes around 6~MHz [Figs.~\ref{figd:displacements}-\ref{fige:displacements}] and 10~MHz [Figs.~\ref{figf:displacements}-\ref{figg:displacements}] are associated to $S_3S_6$ ($f=6.219$~MHz) and $S_5S_{10}$ ($f=10.406$~MHz) in a single Duralumin plate, and are denoted ${S(S_3S_6)^*}$, ${A(S_3S_6)^*}$, ${A(S_5S_{10})^*}$ and ${S(S_5S_{10})^*}$ in the following. The asterisk index indicates that the two modes interacting to create the repulsion differ from the two interacting for lower bonding (see Sec.~\ref{sec:influence_d}) but the displacements are very similar in both cases. 

\section{\label{sec:params}Parametric Study}

In order to understand the behavior of the ZGV mode frequencies, a parametric study is conducted for the same symmetric tri-layer. The Duralumin plate thicknesses and the elastic parameters of all layers are maintained constant, while the epoxy layer thickness and the interfacial stiffnesses are varied successively.

\subsection{\label{sec:influence_d}Influence of the bonding layer thickness} 

The influence of the layer thickness is studied for a strong coupling [$K_L=K_T=10^{17}$~N/m$^3$, Fig.~\ref{figa:EvolZGV_d_KLKT_1e5}] and for a weak coupling [$K_L=K_T=10^{14}$~N/m$^3$, Fig.~\ref{figb:EvolZGV_d_KLKT_1e5}]. Several ZGV modes appear in both cases. Their existence is limited to a thickness range, except for the ZGV mode associated to the lowest frequency ($f_{S_1S_2}$), that exists for all thicknesses. For a layer thickness above $\sim$20~\textmu m, modes ${S(S_1S_2)}$ and ${A(S_1S_2)}$ are present at frequencies around 2~MHz. Modes $S(S_3S_6)$ and $A(S_3S_6)$ around 6~MHz and modes $S(S_5S_{10})$ and $A(S_5S_{10})$ around 10~MHz are also present although for some values of $d$, one or two of these modes is absent. In Figs.~\ref{figa:EvolZGV_d_KLKT_1e5}-\ref{figb:EvolZGV_d_KLKT_1e5}, around 6 and 10~MHz, a ZGV branch is disappearing and another branch is then appearing at a close frequency. Simulations show that displacements associated to those ZGV branches are similar. Thus, the same notation is used for these modes and an asterisk indicates the branch change. It could be noted that for a very thin intermediate layer ($h\leq10$~\textmu m) and a strong bonding, four other modes appear [Fig.~\ref{figa:EvolZGV_d_KLKT_1e5}]; they are not observed for lower spring stiffnesses [Fig.~\ref{figb:EvolZGV_d_KLKT_1e5}]. These modes correspond to ZGV modes in a single plate having twice the thickness of the duralumin plates. For example, frequencies and displacements of the first two additional modes correspond to $S_3S_6$ and $S_5S_{10}$ ZGV modes of a 3~mm thick duralumin plate, respectively.\\

As expected, all ZGV frequencies decrease with increasing bonding layer thickness. However, frequency variations differ from one mode to the other. The ZGV frequency $f_{S_1S_2}$ decays by 26.5\% for $K_L=K_T=10^{17}$~N/m$^3$ [Fig.~\ref{fig:EvolZGV_d_KLKT_1e5}(a)]. This highlights the necessity of an accurate knowledge of the value of each layer thickness before considering the interfacial stiffnesses. Other layer properties involved in the ZGV calculation ($\rho_2$, $V_{L2}$, $V_{T2}$) also influence the ZGV Lamb mode frequencies. However, such properties can be measured before the bonding, and are assumed to be homogeneous. On the contrary, the glue layer thickness can locally evolve and simulations demonstrate the need to precisely determine the layer thickness.

\begin{figure*}[ht!]
\centering
\begin{minipage}[c]{0.7\textwidth}
\subfigure{\includegraphics[width=0.98\textwidth]{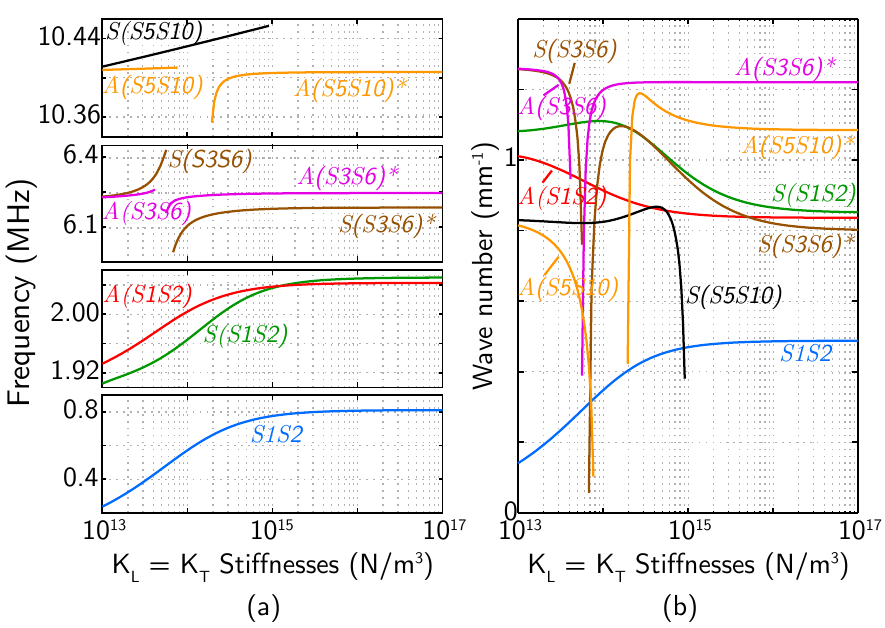}\label{figa:EvolZGV_KLKT_1e1_1e5}}
\subfigure{\label{figb:EvolZGV_KLKT_1e1_1e5}}
\end{minipage}\hfill
\begin{minipage}[c]{0.25\textwidth}
\caption{Theoretical ZGV mode frequencies~(a) and associated wave numbers~(b), for $d=50$~\textmu m, as a function of stiffnesses $K_L=K_T$ varying from $10^{13}$~N/m$^3$ to $10^{17}$~N/m$^3$. Symmetrical modes (solid line) and anti-symmetrical modes (dashed line).}
\label{fig:EvolZGV_KLKT_1e1_1e5}
\end{minipage}
\end{figure*}

\begin{figure*}[!ht]
\centering
\begin{minipage}[c]{0.7\textwidth}
\subfigure{\includegraphics[width=0.98\textwidth]{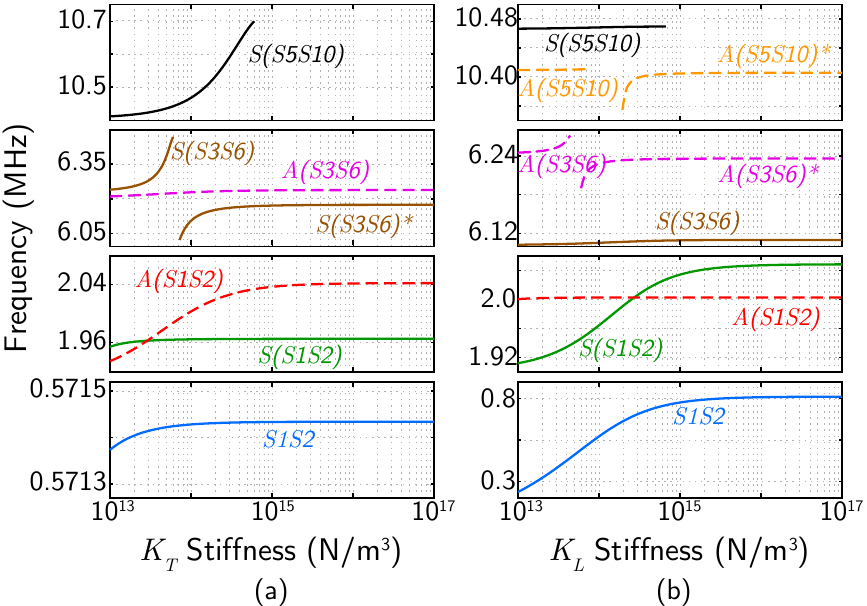}\label{figa:EvolZGV_KL_fixe_KT_evol_KT_fixe_KL_evol}}
\subfigure{\label{figb:EvolZGV_KL_fixe_KT_evol_KT_fixe_KL_evol}}
\end{minipage}\hfill
\begin{minipage}[c]{0.25\textwidth}
\caption{Theoretical ZGV modes of the tri-layer structure with $d=50$~\textmu m as a function of (a)~the stiffness $K_T$ evolving from $10^{13}$~N/m$^3$ to $10^{17}$~N/m$^3$ with $K_L=10^{14}$~N/m$^3$ and of (b)~the stiffness $K_L$ evolving from $10^{13}$~N/m$^3$ to $10^{17}$~N/m$^3$ with $K_T=10^{14}$~N/m$^3$. Symmetrical modes (solid line) and anti-symmetrical modes (dashed line). }
\label{fig:EvolZGV_KL_fixe_KT_evol_KT_fixe_KL_evol}
\end{minipage}
\end{figure*}

\subsection{\label{sec:influence_KL}Influence of longitudinal and shear stiffnesses}

For a bi-layer structure, it was shown that symmetrical modes only vary with the longitudinal interfacial stiffness, while anti-symmetrical ones only vary with the shear interfacial stiffness.~\cite{article-mezil-bci} The situation is more complex for a tri-layer structure where symmetric and anti-symmetric modes depend on both $K_L$ and $K_T$ (Eq.~\ref{eq:matriceS}). To understand the dependence of resonance frequencies with interfacial stiffnesses, we first set the bonding layer thickness to $50$~\textmu m and assume $K_L=K_T$. Figure~\ref{fig:EvolZGV_KLKT_1e1_1e5} displays the evolution of ZGV frequencies for symmetrical and anti-symmetrical modes, as a function of the stiffnesses varying over four decades.\\

It appears that the frequencies of all ZGV modes increase with increasing stiffnesses and that the lowest ZGV mode $S_1S_2$ is, again, always present. As shown for a bi-layer structure, this low frequency mode characterizes the presence of a longitudinal coupling between the two plates.~\cite{article-mezil-bci} The ZGV frequency $f_{S_1S_2}$ exists for any $K_L$ and $K_T$ positive values. Its normal component of the displacement is significant [Fig.~\ref{figa:displacements}] for any $K_L$, $K_T$ values, which allows its optical detection (Sec.~\ref{sec:experiences}). ZGV resonances associated to modes in the intermediate layer also exist but their frequencies are much higher because of the intermediate layer thickness ($d=50$~\textmu m): the first one appears around $22$~MHz. As the mechanical displacement is mainly localized in the bonding layer, these ZGV modes are not experimentally detected and then not discussed.\\

The wavelength of ZGV modes should also be considered to optimize the excitation spot size. For a homogeneous plate, the ratio of the wavelength to the plate thickness depends only on the Poisson ratio,~\cite{article-prada} but in a three-layer structure, it depends on several parameters. For example, Fig.~\ref{figb:EvolZGV_KLKT_1e1_1e5} shows that the wavelength depends on the ZGV mode order and either increases or decreases with the spring stiffnesses $K_{L,T}$. Beginning of a ZGV branch is associated to a low wave number which increases with $K_{L,T}$ [\eg, ${A(S_3S_6)^*}$ in Fig.~\ref{figb:EvolZGV_KLKT_1e1_1e5}]. Similarly, the wave number vanishes at the end of a ZGV branch [\eg, ${A(S_5S_{10})}$ in Fig.~\ref{figb:EvolZGV_KLKT_1e1_1e5}]. The $S_1S_2$ ZGV mode wave number is close to zero for very low stiffnesses and increases with $K_{L,T}$. This behavior can be ascribed to the existence of the ${S_1S_2}$ mode for any strictly positive spring stiffnesses. At the opposite, the existence of other modes in a single plate (\ie, for $K_L=K_T=0$) explains their non-zero wave number for low spring stiffnesses.\\

Figure~\ref{fig:EvolZGV_KL_fixe_KT_evol_KT_fixe_KL_evol} presents the evolution of the first five ZGV mode frequencies as a function of $K_T$ with $K_L=10^{14}$~N/m$^3$ (a) and of $K_L$ with $K_T=10^{14}$~N/m$^3$ (b). For a bi-layer structure, it was demonstrated that symmetrical (anti-symmetrical) ZGV modes only depends on $K_L$ ($K_T$).\cite{article-mezil-bci} For a tri-layer structure, this is no longer true. However, in our experiment the bonding layer thickness is small compared to the plate one and similar behaviors are observed. Therefore, each ZGV mode is mostly influenced by one or the other interfacial stiffness. In particular $f_{S_1S_2}$ and $f_{S(S_1S_2)}$ mainly depend on $K_L$ [Fig.~\ref{figb:EvolZGV_KL_fixe_KT_evol_KT_fixe_KL_evol}], whereas $f_{A(S_1S_2)}$ primarily depends on $K_T$ [Fig.~\ref{figa:EvolZGV_KL_fixe_KT_evol_KT_fixe_KL_evol}]. These observations will guide the experimental procedure to determine the interfacial stiffnesses.

\section{\label{sec:experiences}Experiments}

An all-optical set-up is used to measure ZGV resonances. Lamb waves are excited by a Q-switched Nd:YAG (yttrium aluminium garnet) laser (Quantel Laser, France, Centurion, 1064~nm) which delivers 10-ns pulses of 8-mJ energy. The normal surface displacement is detected on the opposite side of the plate with an interferometer (Bossa Nova Tech., USA, Tempo 1D, 532~nm). The structure is a 100$\times$100~mm$^2$ tri-layer composed of two Duralumin plates of thickness $h=1.50$~mm bonded by a bi-component epoxy adhesive (Loctite Hysol, France, 9466 A\&B). Substrate surfaces were cleaned with acetone. To control the adhesive joint thickness, thin metallic spacers are disposed along the edges prior to the spreading of the glue. The specimen was cured at room temperature ($22\celsius$) and then maintained under a 5-kN constant pressure during 12~hours with an automatic mechanical testing machine (MTS, USA, QTest/25). The thickness spacers were 43-\textmu m thick. However, in the middle of the plate, where the experiment is achieved, the layer thickness is smaller due to the hardening process of the glue. The determined mass densities are $\rho_{1}=2795$~kg/m$^3$ for Duralumin and $\rho_2=1170$~kg/m$^3$ for the epoxy. The longitudinal and shear velocities were measured on a single Duralumin plate using ZGV Lamb modes:~\cite{article-prada-local} $V_{L1}=6398$~m/s and $V_{T1}=3122$~m/s. The longitudinal and shear acoustic wave velocities of epoxy are estimated using piezoelectric contact transducers (Olympus, USA, V103 and V153): $V_{L2}=2500$~m/s and $V_{T2}=1100$~m/s.\\

\begin{figure}[!ht]
\centering
\subfigure{\includegraphics[width=\columnwidth]{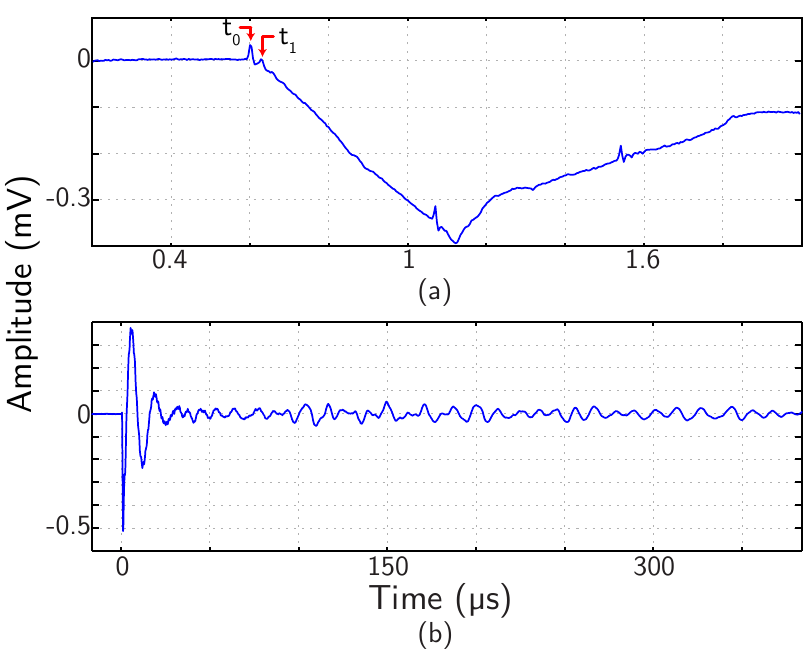}\label{figa:tempscourts}}
\subfigure{\label{figb:tempscourts}}
\caption{Typical signal measured by transmission in the Duralumin/epoxy/Duralumin structure: (a)~at short arrival time with a $\sim100$~µm beam diameter to observe first reflections in the intermediate layer, (b)~with a beam diameter of $\sim3$~mm to enhance ZGV~mode resonances.}
\label{fig:tempscourts}
\end{figure}

For an accurate knowledge of the bonding layer thickness, the precursors in the transmitted signal are obtained with a small laser beam diameter ($\sim100$~µm at $1/e$) [Fig.~\ref{figa:tempscourts}]. The first arrival, at $t_0$, corresponds to the direct transmission in the structure while the second arrival, at $t_1$, is associated with the wave twice reflected in the intermediate layer. The measured times $t_0=601.6\pm0.4$~ns and $t_1=629.6\pm0.4$~ns lead to a glue layer thickness $d=35\pm1$~\textmu m.\\

\begin{figure}[t]
\centering
\includegraphics[width=\columnwidth]{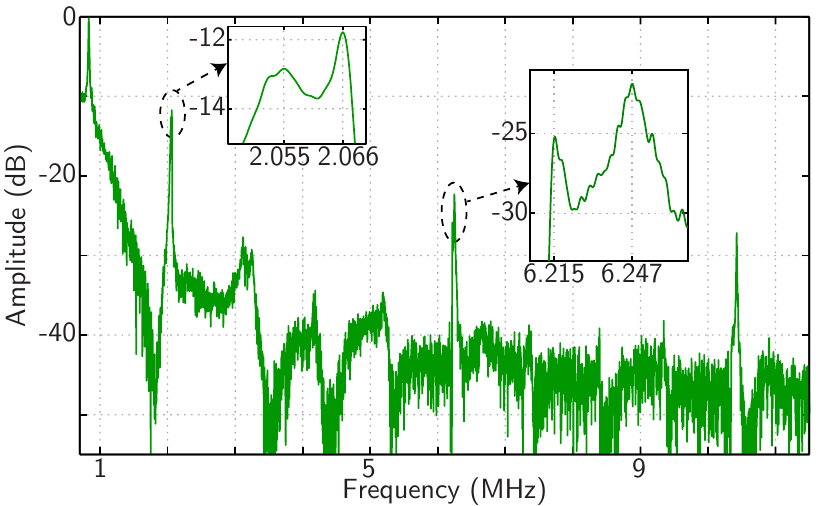}
\caption{Experimental signal obtained by transmission in the Duralumin/epoxy/Duralumin tri-layer structure.}
\label{fig:expZGV}
\end{figure}

The local resonance spectrum is measured at the same point with an enlarged spot. A $\sim3$-mm beam diameter (at $1/e$) is chosen as a good compromise to excite the first ZGV modes [Fig.~\ref{figb:EvolZGV_KLKT_1e1_1e5} and Balogun \etal \cite{article-prada}]. The resulting signal and its Fourier transform are displayed in Fig.~\ref{figb:tempscourts} and Fig.~\ref{fig:expZGV}, respectively. Several ZGV resonances are observed with a precision estimated to be 1~kHz [see Tab.~\ref{table:ZGVfreq}]. The first one at $f=0.836$~MHz, corresponds to the ${S_1S_2}$ mode. The following two around 2.1 MHz correspond to ${S(S_1S_2)}$ and ${A(S_1S_2)}$, not yet discriminated. Similarly, two resonances around 6.2~MHz can be attributed to ${S(S_3S_6)}$ and ${A(S_3S_6)}$. Around 10~MHz, only one resonance is detected. This is probably due to the fact that the displacement associated to the $S(S_5S_{10})$ ZGV mode is mainly localized in the epoxy layer and not at the free surface [see Fig.~\ref{figg:displacements}].\\

\begin{figure}[b]
\centering
\subfigure{\includegraphics[width=\columnwidth]{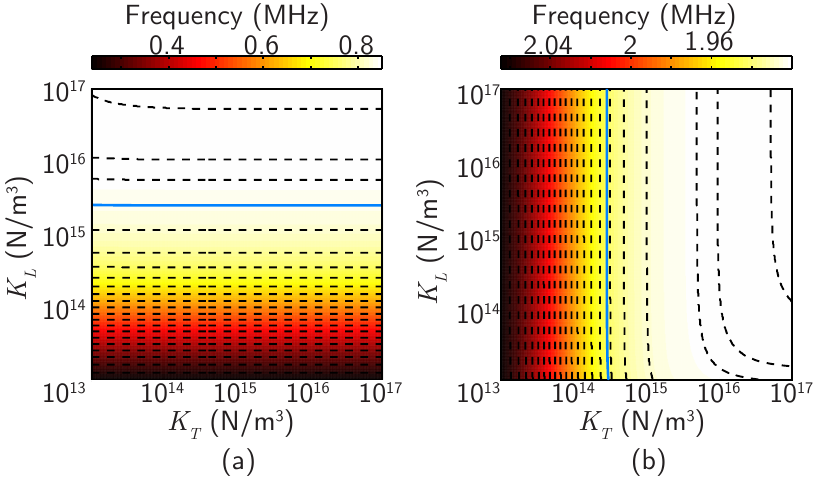}\label{figa:mapZGV}}
\subfigure{\label{figb:mapZGV}}
\caption{Theoretical evolution of the frequency of the modes ${S_1S_2}$~(a) and ${A(S_1S_2)}$~(b) as a function of both interfacial stiffnesses in the range $[10^{13};10^{17}]$~N/m$^3$. Dashed lines correspond to isofrequency curves. Solid lines correspond to $f_{S_1S_2}=0.836$~MHz~(a) and $f_{A(S_1S_2)}=2.055$~MHz~(b). }
\label{fig:mapZGV}
\end{figure}

In order to determine the interfacial stiffnesses, the frequency $f_{S_1S_2}$ and $f_{A(S_1S_2)}$ are calculated as a function of both spring stiffnesses in the range $[10^{13};10^{17}]$~N/m$^3$. The resulting maps are displayed in Figs.~\ref{figa:mapZGV} an \ref{figb:mapZGV}. From the isofrequency curves, it clearly appears that $f_{S_1S_2}$ mostly depends on $K_L$ while $f_{A(S_1S_2)}$ mostly depends on $K_T$. In the experiment, it is necessary to figure out which one of the two resonances measured around 2~MHz corresponds to the anti-symmetrical mode ${A(S_1S_2)}$.\\

The discrimination can be done from the resonance width. As shown in Fig.~\ref{figc:displacements} for the anti-symmetrical mode, the glue layer is mainly sheared, leading to a stronger attenuation than for the symmetrical ones, where the layer is weakly sheared [Fig.~\ref{figb:displacements}]. Consequently, the broader peak at 2.055 MHz can be ascribed to the anti-symmetrical mode $A(S_1S_2)$, and the thinner one at 2.066~MHz to the $S(S_1S_2)$ mode [Fig.~\ref{fig:expZGV}]. A similar observation can be done for $S(S_3S_6)$ at 6.215~MHz and $A(S_3S_6)$ at 6.247~MHz. Moreover, if we consider the reverse case where ${A(S_1S_2)}$ resonance would be at $f=2.066$~MHz, the obtained set of $K_L$ and $K_T$ values increases the discrepancies between theoretical and experimental ZGV frequencies. This confirms the identification of the symmetrical and anti-symmetrical ZGV modes. Consequently, the isofrequency curves are plotted at $f=2.055$~MHz for the anti-symmetrical mode $A(S_1S_2)$ [Figs.~\ref{figb:mapZGV}] and at $f=0.836$~MHz for the $S_1S_2$ mode [Fig.~\ref{figa:mapZGV}].\\

The interfacial stiffnesses correspond to the intersection of these two curves and are found to be $K_L=2.52\times10^{15}$ and $K_T=2.83\times 10^{14}$~N/m$^3$ for a thickness $d=35$~\textmu m. For these values, the theoretical frequencies of the other ZGV modes fit the measured peaks with a discrepancy below $0.3$~\% [Tab.~\ref{table:ZGVfreq}]. For $d$ varying in the range 34 to 36~\textmu m, which corresponds to the layer thickness uncertainty, $K_L$ varies from $2.15$ to $3.07\times 10^{15}$~N/m$^3$ and $K_T$ varies from 2.56 to $3.17\times 10^{14}$~N/m$^3$.\\ 

\begin{table}[t]
\begin{minipage}[c]{\columnwidth}
\begin{ruledtabular}
\centering
\caption{Experimental and theoretical ZGV Lamb mode frequencies.}
\label{table:ZGVfreq}
\begin{tabular}{ccc}
 Experiment (MHz) & Theory (MHz) & Difference (kHz) \\
 \hline
 0.836		& 0.836       &  -    \\ 
 2.055      & 2.055       &  -	  \\
 2.066      & 2.063       &  -3   \\
 6.215      & 6.204       & -11   \\
 6.247      & 6.260       &  13   \\
 10.424     & 10.426      &  2    \\
\end{tabular}
\end{ruledtabular}
\end{minipage}
\end{table}

Similar values for the longitudinal stiffness are found in the literature for aluminium/epoxy interface.~\citep{vlasie09, article-her14} The ratio $K_T/K_L$ equal to 0.11 is much lower that the values of 0.7 to 0.8 reported by Cantrell.~\cite{article-cantrell} However this ratio is close to the value 0.16 measured by Guo \etal , in the case of a film deposited on a steel substrate.~\cite{article-guo-achenbach} It should be noted that parameters $K_T$ and $K_L$ depend on many factors during the hardening process of the bonding layer such as humidity or surface roughness.\\

\begin{figure}[b]
\centering
\includegraphics[width=0.7\columnwidth]{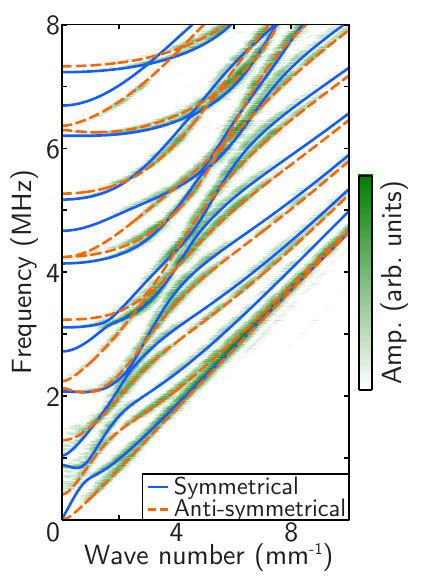}
\caption{Experimental (map) and theoretical (solid and dashed lines) dispersion curves obtained with the measured stiffnesses $K_L = 2.52\times10^{15}$ N/m$^3$ and $K_T = 2.83\times10^{14}$ N/m$^3$.}
\label{fig:DispersionCurves}
\end{figure}

Dispersion curves were measured by shifting the source (diameter spot $0.1$~mm) from the probe beam, along $40$~mm by $0.1$~mm steps, and applying a two-dimensional Fourier transform to the acquired signals. Experimental curves are displayed in Fig.~\ref{fig:DispersionCurves}. Theoretical dispersion curves were calculated with the stiffnesses previously obtained and then superposed to experimental data. A good agreement between theoretical and experimental curves is observed for a large range of $(\omega,k)$. It confirms that, provided the elastic properties and thickness of the epoxy layer are known, ZGV frequencies are sufficient to determine the local interfacial stiffnesses of the structure.

\section{Conclusion}

 A method using Zero-Group Velocity Lamb modes is developed to characterize the bonding in a tri-layer structure. Theoretical and experimental studies are carried out in the symmetrical case of two Duralumin plates bonded by an epoxy layer. The coupling at both epoxy/metal interfaces is modeled by normal and tangential stiffnesses, $K_L$ and $K_T$. The parametric study demonstrates the sensitivity of ZGV Lamb modes to $K_{L,T}$ and to the bonding layer thickness. In general, the first symmetrical and anti-symmetrical ZGV Lamb modes are mostly sensitive to normal and shear stiffnesses, respectively. Provided the bonding layer thickness is well known, the interfacial stiffnesses can be evaluated from ZGV frequency measurements. To this end, the low frequency ZGV Lamb mode is used to estimate $K_L$, and then, the second ZGV Lamb mode is used to estimate $K_T$. Difference between theoretical and the other experimental ZGV frequencies is found to be less than 0.3\% which shows the potential interest of this technique.\\

In the future, measurements with different glue thicknesses or adhesive layers could be investigated to compare different bonding qualities. It would be insightful to couple ZGV interfacial stiffness measurements with mechanical tensile tests.

\begin{acknowledgements}
This work was supported by LABEX WIFI (Laboratory of Excellence ANR-10-LABX-24) within the French Program ``Investments for the Future'' under reference ANR-10-IDEX-0001-02 PSL$^{\ast}$ and by the Chaire SAFRAN ESPCI.
\end{acknowledgements}



\end{document}